\documentclass[a4paper]{llncs}
\usepackage{longtable}
\usepackage{proof}
\usepackage[table]{xcolor}
\usepackage{amsmath}
\usepackage{mathtools}
\usepackage{setspace}
\usepackage{cite}
\usepackage{xcolor}
\usepackage{float}
\usepackage{enumitem}
\usepackage{calligra}
\usepackage{graphicx}
\usepackage{algorithm,algpseudocode}
\usepackage{scalerel}
\usepackage{multirow}
\usepackage{color}

\makeatletter
\newcommand{\algmargin}{\the\ALG@thistlm}
\makeatother
\newlength{\whilewidth}
\algdef{SE}[parWHILE]{parWhile}{EndparWhile}[1]
  {\parbox[t]{\dimexpr\linewidth-\algmargin}{%
     \hangindent\whilewidth\strut\algorithmicwhile\ #1\ \algorithmicdo\strut}}{\algorithmicend\ \algorithmicwhile}%
\algnewcommand{\parState}[1]{\State%
  \parbox[t]{\dimexpr\linewidth-\algmargin}{\strut #1\strut}}
  
\DeclareFontShape{T1}{calligra}{m}{n}{<->s*[2.2]callig15}{}
\DeclareMathAlphabet{\matcalligra}{T1}{calligra}{m}{n}

\DeclareMathAlphabet\mathbfcal{OMS}{cmsy}{b}{n}
\let\mathcal\undefined \DeclareMathAlphabet{\mathcal}{OMS}{cmsy}{m}{n}

\newcommand{\flqsr}{\ensuremath{\mathsf{4LQS^R}}}

\newcommand{\dlssx}{\mathcal{DL}\langle \mathsf{4LQS^{R,\!\times}}\rangle(\D)}
\newcommand{\shdlssx}{\mathcal{DL}_{\D}^{4,\!\times}}

\newcommand{\D}{\mathbf{D}}
\newcommand{\sroiqd}{\mathcal{SROIQ}(\D)}

\newcommand{\defAs}{\coloneqq}
\newcommand{\I}{\mathbf{I}}
\newcommand{\Ind}{\mathbf{Ind}}
\newcommand{\C}{\mathbf{C}}

\newcommand{\Ra}{\mathbf{R_A}}
\newcommand{\Rd}{\mathbf{R_D}}
\newcommand{\sym}{\mathsf{Sym}}
\newcommand{\asym}{\mathsf{Asym}}
\newcommand{\refl}{\mathsf{Ref}}
\newcommand{\irref}{\mathsf{Irref}}
\newcommand{\tra}{\mathsf{Tra}}
\newcommand{\fun}{\mathsf{Fun}}

\newcommand{\vipcomment}[1]{}
\newcommand{\pow}{\mathcal{P}}
\newcommand{\var}{\mathtt{Var}}
\newcommand{\T}{\mathcal{T}}

\newcommand{\ke}{KE-tableau}
\newcommand{\M}{\mathbfcal{M}}

\newcommand{\KB}{\mathcal{KB}}
\newcommand{\vari}{\mathtt{Var}_i}
\newcommand{\varz}{\mathtt{Var}_0}

\newcommand{\vardt}{\mathsf{V}_{\mathsf{d}}}

\newcommand{\vare}{\mathsf{V}_{\mathsf{e}}}
\newcommand{\varar}{\mathsf{V}_{\mathsf{ar}}}
\newcommand{\varcon}{\mathsf{V}_{\mathsf{c}}}
\newcommand{\varind}{\mathsf{V}_{\mathsf{i}}}
\newcommand{\varcr}{\mathsf{V}_{\mathsf{cr}}}
\newcommand{\sfvar}[2]{ {\mathsf{#1}_{#2}} }

\newcommand{\DT}{\mathcal{D}}

\newcommand{\coreflqsr}{\mathsf{4LQS}_{\scriptscriptstyle{\mathcal{DL}_{ \scaleto{\mathbf{D}}{2.5pt}}^{\scaleto{4,\!\times}{3pt}}}}^\mathsf{R}}

\newcommand{\keg}{\textnormal{KE}$^{\mathbf{\gamma}}$-tableau} 
\newcommand{\kegx}{\textnormal{KE}$^{\mathbf{\gamma}}$-tableaux} 
\newcommand{\prochoplus}{\textit{HOCQA}^\gamma\textit{-}{\shdlssx}} 
\newcommand{\egamma}{\textnormal{E}^{\gamma}\textnormal{-rule}}
\newcommand{\seqs}{\mathcal{S}^{\overline{\beta}_i\tau}}
\newcommand{\seqsnj}{\mathcal{S}^{\overline{\beta}\tau}_j}
\newcommand{\consistency}{\textit{Consistency-}{\shdlssx}} 
\newcommand{\litqt}{Lit^{\vartheta}_q}
\newcommand{\minord}{<_{x_0}}

\title{A set-based reasoner for the description logic $\shdlssx$ (Extended Version)}

\author{Domenico Cantone \and Marianna Nicolosi-Asmundo \and \\Daniele Francesco Santamaria}
\institute{
University of Catania, Dept. of Mathematics and Computer Science\\
~email:~\texttt{\{cantone,nicolosi,santamaria\}@dmi.unict.it}
}

\graphicspath{{img/}}

\begin{document}
\maketitle


\begin{abstract}
	
	We present a \ke-based implementation of a reasoner for a decidable fragment of (stratified) set theory expressing the description logic $\dlssx$ ($\shdlssx$, for short). Our application solves the main TBox and ABox reasoning problems for $\shdlssx$. In particular, it solves the consistency problem for $\shdlssx$-knowledge bases represented in set-theoretic terms, and a generalization of the \emph{Conjunctive Query Answering} problem in which conjunctive queries with variables of three sorts are admitted. 
The reasoner, which extends and optimizes a previous prototype for the consistency checking of $\shdlssx$-knowledge bases (see \cite{cilc17}), is implemented in \textsf{C++}. It supports $\shdlssx$-knowledge bases serialized in the OWL/XML format, and it admits also rules expressed in SWRL (Semantic Web Rule Language).

\end{abstract}

\section{Introduction}

A wealth of decidability results has been collected over the years within the research field of \emph{Computable Set Theory} \cite{CaFeOm90,CaOmPo01,SchwCanOmoPol11}. However, only recently some of these results have been applied in the context of knowledge representation and reasoning for the semantic web. Such efforts have been motivated by the characteristics of the set-theoretic fragments considered, as they provide very expressive unique formalisms that combine the modelling capabilities of a rule language with the constructs of description logics. The decidable multi-sorted quantified set-theoretic fragment $\flqsr$ \cite{CanNic2013} is appropriate in this sense, in consideration of the fact that its decision procedure is efficiently implementable. We recall that the language of $\flqsr$ involves variables of four sorts, pair terms, and a restricted form of quantification. 




In \cite{RR2017}, the theory $\flqsr$ has been used to represent the expressive description logic $\shdlssx$ by means of a suitable translation mapping. Moreover, decidability of the most widespread reasoning problems for $\shdlssx$, such as the consistency problem and the Conjunctive Query Answering (CQA) problem for $\shdlssx$-knowledge bases (KBs) were proved via a reduction to the satisfiability problem for $\flqsr$. Since $\flqsr$ admits variables of four sorts, the CQA problem was generalized in such a way as to admit queries over three sorts of variables. Such a generalization, called Higher-Order Conjunctive Query Answering (HOCQA) problem can be instantiated to the most widespread reasoning tasks for $\shdlssx$-ABox.

The description logic $\shdlssx$ admits Boolean operators on
concepts and abstract roles, concept domain and range, and existential and minimum cardinality  restriction on the left-hand side of inclusion axioms. It also supports role chains on the left-hand side of inclusion axioms and properties on roles such as transitivity, symmetry, and reflexivity. 
In \cite{ictcs16}, its consistency problem has been shown to be NP-complete under not very restrictive constraints. Such a low complexity result depends on the fact that existential quantification cannot appear on the right-hand side of inclusion axioms. Nonetheless, $\shdlssx$
turns out to be more expressive than other low complexity
logics such as OWL RL \cite{santaLAP} and therefore it is very suitable for representing real-world
ontologies. For instance, the restricted version of $\shdlssx$ mentioned above allows
one to express several OWL ontologies, such as \textsf{ArcheOntology} \cite{santaLAP} and \textsf{OntoCeramic} \cite{cilc15}, for the classification of archaeological finds, and \textsf{ArchivioMuseoFabbrica} \cite{jlis17}, concerning the renovation of the Monastery of San Nicola l'Arena in Catania by the architect Giancarlo De Carlo. Since existential quantification is admitted only on the left-hand side of inclusion axioms, $\shdlssx$ is less expressive than logics such as
$\sroiqd\space$ \cite{Horrocks2006} as long as the generation of new individuals is concerned. On
the other hand, $\shdlssx$ is more liberal than $\sroiqd\space$ in the definition of role
inclusion axioms, as the roles involved in $\shdlssx$ are not subject to any
ordering relationship, and the notion of simple role is not needed. For example,
the role hierarchy presented in \cite[page~2]{Horrocks2006} is not expressible in $\sroiqd\space$,
but can be represented in $\shdlssx$. In addition, $\shdlssx$ is a powerful rule language
able to express rules with negated atoms such as 
\\\centerline{
$\mathit{Person}(?p) \wedge \neg \mathit{hasHome}(?p, ?h) \implies \mathit{HomelessPerson}(?p)$
}
that are not supported by the SWRL language.

In \cite{cilc17}, we presented a first effort to implement in \textsf{C++} a \ke-based decision procedure for the consistency problem of $\shdlssx$-KBs, by resorting to the algorithm introduced in \cite{RR2017}. The choice of \ke\space systems \cite{dagostino94}, instead of traditional semantic tableaux \cite{smullyan1995first}, was motivated by the fact that \ke\space systems introduce an analytic cut rule which permits to construct trees whose branches define mutually exclusive situations, thus avoiding the proliferation of redundant branches, typical of Smullyan's semantic tableaux \cite{smullyan1995first}. Thus, given as input a consistent KB, the procedure yields a \ke\space whose open branches induce distinct models of the KB. Otherwise, a closed \ke\space is returned. 

In this contribution we improve the reasoner presented in \cite{cilc17} by introducing a system called \keg\space  which admits a generalization of the KE-elimination rule incorporating the $\gamma$-rule, namely the expansion rule for handling universally quantified formulae. The reasoner also includes a procedure to compute the HOCQA problem for $\shdlssx$. Finally, through suitable benchmark tests, we show that such a novel reasoner is more efficient than the one introduced in \cite{cilc17}.
\vspace{2cm}

\section{Preliminaries}

\subsection{The set-theoretic fragment} \label{4LQS}

We summarize the set-theoretic notions underpinning the description logic $\shdlssx$ and its reasoning tasks. For the sake of conciseness, we avoid to report here the syntax and semantics of the whole $\flqsr$ theory (the interested reader can find it in \cite{CanNic2013} together with the decision procedure for its satisfiability problem). Thus, we focus on the $\flqsr$-formulae \emph{de facto} involved in the set-theoretic representation of $\shdlssx$, namely propositional combinations of $\flqsr$-literals (atomic formulae or their negations) and $\flqsr$ purely universal formulae of the types displayed in Table \ref{tablecore}.
The class of such $\flqsr$-formulae is called $\coreflqsr$.

We recall that the fragment $\flqsr$ admits four collections, $\var_i$, of variables of sort $i$ denoted by $X^i,Y^i,Z^i, \ldots$, for $i=0,1,2,3$ (variables of sort $0$ are also denoted by $x,y,z, \ldots$). Besides variables, also  \textit{pair terms} of the form $\langle x,y \rangle$, with $x,y \in \var_0$, are allowed. Since the types of formulae displayed in Table \ref{tablecore} do not contain variables of sort $2$, here we limit ourselves only to notions and definitions relative to $\coreflqsr$-formulae 
involving variables of sorts $0,1,$ and $3$. 

\vspace*{-0.5cm}
\begin{table}[]
	\centering	\scriptsize	
	\renewcommand{\arraystretch}{1.8}
	\begin{tabular}{cc}
		\hline
		
		\multicolumn{1}{|c|}{\cellcolor{lightgray} Literals of level 0}      & \multicolumn{1}{c|}{\cellcolor{lightgray}Purely universal quantified formulae of level 1}                 \\ 
		\hline
		\multicolumn{1}{|c|}{$x=y, \; x \in X^1, \; \langle x,y \rangle \in X^3$} & \multicolumn{1}{c|}{\multirow{2}{*}{ \shortstack[c]{$(\forall z_1)\ldots(\forall z_n)\varphi_0$, where $z_1, \ldots, z_n \in \varz $ and $\varphi_0$ is \\any propositional combination of \\literals of level 0.   }}} \\
		\multicolumn{1}{|c|}{$\neg (x=y), \; \neg(x \in X^1), \; \neg(\langle x,y \rangle \in X^3)$} & 	
		
		\multicolumn{1}{c|}{}                       \\
		\cline{1-2}
		&                                             \\
	\end{tabular}
	
	\caption{Types of literals and quantified formulae admitted in $\coreflqsr$.}
	\label{tablecore}
\end{table}
\vspace*{-0.5cm}

The variables $z_1,\ldots,z_n$ are said to occur \textit{quantified} in $(\forall z_1) \ldots (\forall z_n) \varphi_0$. A variable occurs \textit{free} in a $\coreflqsr$-formula $\varphi$ if it does not occur quantified in any subformula of $\varphi$. For $i = 0,1,3$, we denote with $\vari(\varphi)$ the collections of variables of sort $i$ occurring free in $\varphi$.


Given sequences of distinct variables $\vec{x}$ (in $\var_0$), $\vec{X}^{1}$ (in $\var_1$), and $\vec{X}^{3}$ (in $\var_3$), of length $n$, $m$, and $q$, respectively, and sequences of (not necessarily distinct) variables $\vec{y}$ (in $\var_0$), $\vec{Y}^{1}$ (in $\var_1$), and $\vec{Y}^{3}$ (in $\var_3$), also of length $n$, $m$, and $q$, respectively, the 
$\coreflqsr$-substitution $\sigma \defAs \{ \vec{x}/\vec{y}, \vec{X}^1/\vec{Y}^1, \vec{X}^3/\vec{Y}^3\}$ is the mapping $\varphi \mapsto \varphi\sigma$ such that, for any given universal quantified $\coreflqsr$-formula $\varphi$, $\varphi\sigma$ is the result of replacing in $\varphi$ the free occurrences of the variables $x_i$ in $\vec{x}$ (for $i = 1,\ldots, n$)  
with the corresponding $y_i$ in $\vec{y}$, of $X^1_j$ in $\vec{X}^1$ (for $j = 1,\ldots,m$) with $Y^1_j$ in $\vec{Y}^1$, and of $X^3_h$ in $\vec{X}^3$ (for $h= 1,\ldots,q$) with $Y^3_h$ in $\vec{Y}^3$, respectively.  A substitution $\sigma$ is \emph{free} for $\varphi$ if the formulae $\varphi$ and $\varphi\sigma$ have exactly the same 
occurrences of quantified variables. The \emph{empty substitution}, denoted $\epsilon$, satisfies $\varphi \epsilon = \varphi$, for each $\coreflqsr$-formula $\varphi$.

A $\coreflqsr$-\emph{interpretation} is a pair $\mathbfcal{M}=(D,M)$, where $D$ is a nonempty collection of objects (called \emph{domain} or \emph{universe} of $\mathbfcal{M}$) and $M$ is an assignment over the variables in $\mathtt{Var}_i$, for $i=0,1,3$,  such that:\\\centerline{$MX^{0} \in D,MX^1 \in \pow(D),  MX^3 \in \pow(\pow(\pow(D))),$} where $ X^{i} \in \mathtt{Var}_i$, for $i=0,1,3$, and $\pow(s)$ denotes the powerset of $s$.

\smallskip
\noindent
Pair terms are interpreted \emph{\`a la} Kuratowski, and therefore we put \\[.1cm]
\centerline{$M \langle x,y \rangle \defAs \{ \{ Mx \},\{ Mx,My \} \}$.}
Next, let
\begin{itemize}[topsep=0.1cm, itemsep=0cm]
	\item[-] $\mathbfcal{M}=(D,M)$ be a $\coreflqsr$-interpretation,
	
	\item[-] $x_1,\ldots,x_n \in \mathtt{Var}_0$, and
	
	\item[-] $u_1, \ldots, u_n \in D$.
\end{itemize}

\smallskip
\noindent
By $\mathbfcal{M}[ \vec{x}  / \vec{u}]$, we denote the interpretation $\mathbfcal{M}'=(D,M')$ such that $M'x_i =u_i$ (for $i=1,\ldots,n$), and which otherwise coincides with $M$ on all remaining variables. For a $\coreflqsr$-interpretation $\mathbfcal{M} =(D,M)$ and a formula $\varphi$, the satisfiability relationship $ \mathbfcal{M} \models \varphi$ is recursively defined over the structure of $\varphi$ as follows. Literals are evaluated in a standard way, based on the usual interpretation of the predicates `$\in$'
and `$=$', and of the propositional negation 
 `$\neg$'. Compound formulae are interpreted according to the standard rules of propositional logic. Finally, purely universal formulae are evaluated as follows:
 
\begin{itemize}[topsep=0.1cm, itemsep=0.1cm]
	\item[-] $\mathbfcal{M}  \models (\forall z_1) \ldots (\forall z_n) \varphi _0$ iff $\mathbfcal{M}   [ \vec{z} / \vec{u}] \models \varphi_0$, for all $\vec{u} \in D^{n}$. 
\end{itemize}

If $\mathbfcal{M} \models \varphi$, then $\mathbfcal{M} $ is said to be a $\coreflqsr$-model for $\varphi$. A $\coreflqsr$-formula is said to be \emph{satisfiable} if it has a $\coreflqsr$-model. A $\coreflqsr$-formula is \emph{valid} if it is satisfied by all $\coreflqsr$-interpretations. 

\subsection{The logic $\dlssx$}\label{dlssx}
It is convenient to recall the main notions and definitions concerning the description logic $\dlssx$ (also called $\shdlssx$) \cite{ictcs16}.

Let $\Ra$, $\Rd$, $\mathbf{C}$, and $\Ind$ be denumerable pairwise disjoint sets of abstract role names, concrete role names, concept names, and individual names, respectively. We assume that the set of abstract role names $\Ra$ contains a name $U$ denoting the universal role. 

Data types are introduced through the notion of data type maps, defined  according to \cite{Motik2008} as follows. A \emph{data type map} is a quadruple $\D = (N_{D}, N_{C},N_{F},\cdot^{\D})$, where  $N_{D}$ is a finite set of data types, $N_{C}$ is a function assigning a set of constants $N_{C}(d)$ to each data type $d \in N_{D}$, $N_{F}$ is a function assigning a set of facets $N_{F}(d)$ to each $d \in N_{D}$, and $\cdot^{\D}$ is (i) a function assigning a data type interpretation $d^{\D}$ to each data type $d \in N_{D}$, (ii) a facet interpretation $f^{\D} \subseteq d^{\D}$ to each facet $f \in N_{F}(d)$, and (iii) a data value $e_{d}^{\D} \in d^{\D}$ to every constant $e_{d} \in N_{C}(d)$. Facets determine subsets of data values considered of interest in a specific application domain. We shall assume that the interpretations of the data types in $N_{D}$ are nonempty pairwise disjoint sets.

%
%
%
%

%

\vipcomment{An abstract role hierarchy $\mathsf{R}_{a}^{H}$ is a finite collection of RIAs.  A strict partial order $\prec$ on  $\Ra \cup \{ R^- \mid R \in \Ra \}$ is called \emph{a regular order} if $\prec$ satisfies, additionally, $S \prec R$ iff $S^- \prec R$, for all roles R and S.\footnote{We recall that a strict partial order $\prec$  on a set $A$ is an irreflexive and transitive relation on $A$.}}

\noindent
(a) \emph{$\shdlssx$-data types}, (b) \emph{$\shdlssx$-concepts}, (c) \emph{$\shdlssx$-abstract roles}, and (d) \emph{$\shdlssx$-concrete role terms} are defined according to the DL standard notation (see \cite{Horrocks2006}) as follows:
\begin{itemize}
	\item[(a)] $t_1, t_2 \longrightarrow dr ~|~\neg t_1 ~|~t_1 \sqcap t_2 ~|~t_1 \sqcup t_2 ~|~\{e_{d}\}\,,$
	
	\item[(b)] $C_1, C_ 2 \longrightarrow A ~|~\top ~|~\bot ~|~\neg C_1 ~|~C_1 \sqcup C_2 ~|~C_1 \sqcap C_2 ~|~\{a\} ~|~\exists R.\mathit{Self}| \exists R.\{a\}| \exists P.\{e_{d}\}\, ,$
	
	\item[(c)] $R_1, R_2 \longrightarrow S ~|~U ~|~R_1^{-} ~|~ \neg R_1 ~|~R_1 \sqcup R_2 ~|~R_1 \sqcap R_2 ~|~R_{C_1 |} ~|~R_{|C_1} ~|~R_{C_1 ~|~C_2} ~|~id(C) ~|~ $
	
	$C_1 \times C_2    \, ,$
	
	\item[(d)] $P_1,P_2 \longrightarrow T ~|~\neg P_1 ~|~ P_1 \sqcup P_2 ~|~ P_1 \sqcap P_2  ~|~P_{C_1 |} ~|~P_{|t_1} ~|~P_{C_1 | t_1}\, ,$
\end{itemize}
where $dr$ is a data range for $\D$, $t_1,t_2$ are data type terms, $e_{d}$ is a constant in $N_{C}(d)$,  $a$ is an individual name, $A$ is a concept name, $C_1, C_2$ are $\shdlssx$-concept terms, $S$ is an abstract role name,  $R, R_1,R_2$ are $\shdlssx$-abstract role terms, $T$ is a concrete role name, and $P,P_1,P_2$ are $\shdlssx$-concrete role terms. Notice that data type terms are intended to represent derived data types.

A \emph{$\shdlssx$-KB} is a triple ${\mathcal K} = (\mathcal{R}, \mathcal{T}, \mathcal{A})$ such that $\mathcal{R}$ is a $\shdlssx$-RBox, $\mathcal{T}$ is a $\shdlssx$-TBox, and $\mathcal{A}$ a $\shdlssx$-ABox. 

A $\shdlssx$-$RBox$ is a collection of statements of the following types: 
\[
\begin{array}{cccccccccc}
R_1 \equiv R_2,~&R_1 \sqsubseteq R_2,~&~~R_1\ldots R_n \sqsubseteq R_{n+1},~&\sym(R_1),~&\asym(R_1),\\
\refl(R_1),~& \irref(R_1),~&\mathsf{Dis}(R_1,R_2),~&\tra(R_1),~&\fun(R_1),\\
R_1 \equiv C_1 \times C_2,~&P_1 \equiv P_2,~&P_1 \sqsubseteq P_2,~&\mathsf{Dis}(P_1,P_2),~&\fun(P_1),
\end{array}
\]
where $R_1,R_2$ are $\shdlssx$-abstract role terms, $C_1, C_2$ are $\shdlssx$-abstract concept terms, and $P_1,P_2$ are $\shdlssx$-concrete role terms. Any expression of the type $R_1\ldots R_n \sqsubseteq R$, where $R_1,\ldots,R_n,R$ are $\shdlssx$-abstract role terms, is called a \emph{role inclusion axiom (RIA)}. 

A $\shdlssx$-$TBox$ is a set of statements of the types:
\begin{itemize}
	\item[-] $C_1 \equiv C_2$, $C_1 \sqsubseteq C_2$, $C_1 \sqsubseteq \forall R.C_2$, $\exists R.C_1 \sqsubseteq C_2$, $\geq_n\!\! R. C_1 \sqsubseteq C_2$, $C_1 \sqsubseteq {\leq_n\!\! R. C_2}$,
	\item[-] $t_1 \equiv t_2$,~ $t_1 \sqsubseteq t_2$,~ $C_1 \sqsubseteq \forall P.t_1$,~~ $\exists P.t_1 \sqsubseteq C_1$,~~ $\geq_n\!\! P. t_1 \sqsubseteq C_1$,~~ $C_1 \sqsubseteq {\leq_n\!\! P. t_1}$,
\end{itemize}
where $C_1,C_2$ are $\shdlssx$-concept terms, $t_1,t_2$ data type terms, $R$  a $\shdlssx$-abstract role term, and $P$ a $\shdlssx$-concrete role term. Statements of the form $C \sqsubseteq D$, where  $C$ and $D$ are $\shdlssx$-concept terms, are  
\emph{general concept inclusion axioms}.

A $\shdlssx$-$ABox$ is a set of \emph{individual assertions} of the forms: 
\[a : C_1,~~(a,b) : R_1,~~~~ 
a=b,~~~~a \neq b,~~~~e_{d} : t_1,~~~~(a, e_{d}) : P_1,
\] 
with $C_1$ a $\shdlssx$-concept term, $d$ a data type, $t_1$ a data type term, $R_1$ a $\shdlssx$-abstract role term, $P_1$ a $\shdlssx$-concrete role term, $a,b$ individual names, and $e_{d}$ a constant in $N_{C}(d)$.

The semantics of $\shdlssx$ is given via interpretations of the form $\I= (\Delta^\I, \Delta_{\D}, \cdot^\I)$, where $\Delta^\I$ and $\Delta_{\D}$ are nonempty disjoint domains such that $d^\D\subseteq \Delta_{\D}$, for every $d \in N_{D}$, and
$\cdot^\I$ is an interpretation function.
The interpretation of concepts and roles, axioms and assertions is defined in  Table \ref{semdlss}.

{\small
	\begin{longtable}{|>{\centering}m{2.5cm}|c|>{\centering\arraybackslash}m{6.7cm}|}
		\hline
		Name & Syntax & Semantics \\
		\hline
		
		concept & $A$ & $ A^\I \subseteq \Delta^\I$\\
		
		ab. (resp., cn.) rl. & $R$ (resp., $P$ )& $R^\I \subseteq \Delta^\I \times \Delta^\I$ \hspace*{0.5cm} (resp., $P^\I \subseteq \Delta^\I \times \Delta_\D$)\\
		
		
		individual& $a$& $a^\I \in \Delta^\I$\\
		
		nominal & $\{a\}$ & $\{a\}^\I = \{a^\I \}$\\
		
		dtype  (resp., ng.) & $d$ (resp., $\neg d$)& $ d^\D \subseteq \Delta_\D$ (resp., $\Delta_\D \setminus d^\D $)\\
		
		
		negative data type term & $ \neg t_1 $ & $  (\neg t_1)^{\D} = \Delta_{\D} \setminus t_1^{\D}$ \\
		
		data type terms intersection & $ t_1 \sqcap t_2 $ & $  (t_1 \sqcap t_2)^{\D} = t_1^{\D} \cap t_2^{\D} $ \\
		
		data type terms union & $ t_1 \sqcup t_2 $ & $  (t_1 \sqcup t_2)^{\D} = t_1^{\D} \cup t_2^{\D} $ \\
		
		constant in $N_{C}(d)$ & $ e_{d} $ & $ e_{d}^\D \in d^\D$ \\
		
		
		
		\hline
		data range  & $\{ e_{d_1}, \ldots , e_{d_n} \}$& $\{ e_{d_1}, \ldots , e_{d_n} \}^\D = \{e_{d_1}^\D \} \cup \ldots \cup \{e_{d_n}^\D \} $ \\
		
		data range   &  $\psi_d$ & $\psi_d^\D$\\
		
		data range    & $\neg dr$ &  $\Delta_\D \setminus dr^\D $\\
		
		\hline
		
		top (resp., bot.) & $\top$ (resp., $\bot$ )& $\Delta^\I$  (resp., $\emptyset$)\\
		
		
		negation & $\neg C$ & $(\neg C)^\I = \Delta^\I \setminus C$ \\
		
		conj. (resp., disj.) & $C \sqcap D$ (resp., $C \sqcup D$)& $ (C \sqcap D)^\I = C^\I \cap D^\I$  (resp., $ (C \sqcup D)^\I = C^\I \cup D^\I$)\\
		
		
		valued exist. quantification & $\exists R.{a}$ & $(\exists R.{a})^\I = \{ x \in \Delta^\I : \langle x,a^\I \rangle \in R^\I  \}$ \\
		
		data typed exist. quantif. & $\exists P.{e_{d}}$ & $(\exists P.e_{d})^\I = \{ x \in \Delta^\I : \langle x, e^\D_{d} \rangle \in P^\I  \}$ \\

		self concept & $\exists R.\mathit{Self}$ & $(\exists R.\mathit{Self})^\I = \{ x \in \Delta^\I : \langle x,x \rangle \in R^\I  \}$ \\
		
		nominals & $\{ a_1, \ldots , a_n \}$& $\{ a_1, \ldots , a_n \}^\I = \{a_1^\I \} \cup \ldots \cup \{a_n^\I \} $ \\
		
		\hline
		
		universal role & U & $(U)^\I = \Delta^\I \times \Delta^\I$\\
		
		inverse role & $R^-$ & $(R^-)^\I = \{\langle y,x \rangle  \mid \langle x,y \rangle \in R^\I\}$\\
		
		concept cart. prod. & $ C_1 \times C_2$   &  $ (C_1 \times C_2)^I = C_1^I \times C_2^I$ \\
		
		abstract role complement & $ \neg R $ & $ (\neg R)^\I=(\Delta^\I \times \Delta^\I) \setminus R^\I $\\
		
		abstract role union & $R_1 \sqcup R_2$ & $ (R_1 \sqcup R_2)^\I = R_1^\I \cup R_2^\I $\\
		
		abstract role intersection & $R_1 \sqcap R_2$ & $ (R_1 \sqcap R_2)^\I = R_1^\I \cap R_2^\I $\\
		
		abstract role domain restr. & $R_{C \mid }$ & $ (R_{C \mid })^\I = \{ \langle x,y \rangle \in R^\I : x \in C^\I  \} $\\

		concrete role complement & $ \neg P $ & $ (\neg P)^\I=(\Delta^\I \times \Delta^\D) \setminus P^\I $\\
		
		concrete role union & $P_1 \sqcup P_2$ & $ (P_1 \sqcup P_2)^\I = P_1^\I \cup P_2^\I $\\
		
		concrete role intersection & $P_1 \sqcap P_2$ & $ (P_1 \sqcap P_2)^\I = P_1^\I \cap P_2^\I $\\
		
		concrete role domain restr. & $P_{C \mid }$ & $ (P_{C \mid })^\I = \{ \langle x,y \rangle \in P^\I : x \in C^\I  \} $\\
		
		concrete role range restr. & $P_{ \mid t}$ &  $ (P_{\mid t})^\I = \{ \langle x,y \rangle \in P^\I : y \in t^\D  \} $\\
		
		concrete role restriction & $P_{ C_1 \mid t}$ &  $ (P_{C_1 \mid t})^\I = \{ \langle x,y \rangle \in P^\I : x \in C_1^\I \wedge y \in t^\D  \} $\\
		
		\hline
		
		concept subsum. & $C_1 \sqsubseteq C_2$ & $\I \models_\D C_1 \sqsubseteq C_2 \; \Longleftrightarrow \; C_1^\I \subseteq C_2^\I$ \\
		
		ab. role subsum. & $ R_1 \sqsubseteq R_2$ & $\I \models_\D R_1 \sqsubseteq R_2 \; \Longleftrightarrow \; R_1^\I \subseteq R_2^\I$\\
		
		role incl. axiom & $R_1 \ldots R_n \sqsubseteq R$ & $\I \models_\D R_1 \ldots R_n \sqsubseteq R  \; \Longleftrightarrow \; R_1^\I\circ \ldots \circ R_n^\I \subseteq R^\I$\\
		cn. role subsum. & $ P_1 \sqsubseteq P_2$ & $\I \models_\D P_1 \sqsubseteq P_2 \; \Longleftrightarrow \; P_1^\I \subseteq P_2^\I$\\
		
		\hline
		
		symmetric role & $\sym(R)$ & $\I \models_\D \sym(R) \; \Longleftrightarrow \; (R^-)^\I \subseteq R^\I$\\
		
		asymmetric role & $\asym(R)$ & $\I \models_\D \asym(R) \; \Longleftrightarrow \; R^\I \cap (R^-)^\I = \emptyset $\\
		
		transitive role & $\tra(R)$ & $\I \models_\D \tra(R) \; \Longleftrightarrow \; R^\I \circ R^\I \subseteq R^\I$\\
		
		disj. ab. role & $\mathsf{Dis}(R_1,R_2)$ & $\I \models_\D \mathsf{Dis}(R_1,R_2) \; \Longleftrightarrow \; R_1^\I \cap R_2^\I = \emptyset$\\
		
		reflexive role & $\refl(R)$& $\I \models_\D \refl(R) \; \Longleftrightarrow \; \{ \langle x,x \rangle \mid x \in \Delta^\I\} \subseteq R^\I$\\
		
		irreflexive role & $\irref(R)$& $\I \models_\D \irref(R) \; \Longleftrightarrow \; R^\I \cap \{ \langle x,x \rangle \mid x \in \Delta^\I\} = \emptyset  $\\
		
		func. ab. role & $\fun(R)$ & $\I \models_\D \fun(R) \; \Longleftrightarrow \; (R^{-})^\I \circ R^\I \subseteq  \{ \langle x,x \rangle \mid x \in \Delta^\I\}$  \\
		
		disj. cn. role & $\mathsf{Dis}(P_1,P_2)$ & $\I \models_\D \mathsf{Dis}(P_1,P_2) \; \Longleftrightarrow \; P_1^\I \cap P_2^\I = \emptyset$\\
		
		func. cn. role & $\fun(P)$ & $\I \models_\D \fun(p) \; \Longleftrightarrow \; \langle x,y \rangle \in P^\I \mbox{ and } \langle x,z \rangle \in P^\I \mbox{ imply } y = z$  \\
		
		\hline
		
		data type terms equivalence & $ t_1 \equiv t_2 $ & $ \I \models_{\D} t_1 \equiv t_2 \Longleftrightarrow t_1^{\D} = t_2^{\D}$\\
		
		data type terms diseq. & $ t_1 \not\equiv t_2 $ & $ \I \models_{\D} t_1 \not\equiv t_2 \Longleftrightarrow t_1^{\D} \neq t_2^{\D}$\\
		
		data type terms subsum. & $ t_1 \sqsubseteq t_2 $ &  $ \I \models_{\D} (t_1 \sqsubseteq t_2) \Longleftrightarrow t_1^{\D} \subseteq t_2^{\D} $ \\
		
		\hline
		
		concept assertion & $a : C_1$ & $\I \models_\D a : C_1 \; \Longleftrightarrow \; (a^\I \in C_1^\I) $ \\
		
		agreement & $a=b$ & $\I \models_\D a=b \; \Longleftrightarrow \; a^\I=b^\I$\\
		
		disagreement & $a \neq b$ & $\I \models_\D a \neq b  \; \Longleftrightarrow \; \neg (a^\I = b^\I)$\\
		
		
		ab. role asser. & $ (a,b) : R $ & $\I \models_\D (a,b) : R \; \Longleftrightarrow \;  \langle a^\I , b^\I \rangle \in R^\I$ \\
		
		cn. role asser. & $ (a,e_d) : P $ & $\I \models_\D (a,e_d) : P \; \Longleftrightarrow \;   \langle a^\I , e_d^\D \rangle \in P^\I$ \\

		\hline \caption{Semantics of $\shdlssx$.}\\
		\caption*{\emph{Legenda.} \emph{ab.}: abstract, \emph{cn.}: concrete, \emph{rl.}: role, \emph{ind.}: individual, \emph{d. cs.}: data type constant, \emph{dtype}: data type, \emph{ng.}: negated, \emph{bot.}: bottom, \emph{incl.}: inclusion, \emph{asser.}: assertion.}  \label{semdlss}
\end{longtable}}


Let $\mathcal{R}$, $\mathcal{T}$, and $\mathcal{A}$  be as above. An interpretation $\I= (\Delta ^ \I, \Delta_{\D}, \cdot ^ \I)$ is a $\D$-model of $\mathcal{R}$ (resp., $\mathcal{T}$), and we write $\I \models_{\D} \mathcal{R}$ (resp., $\I \models_{\D} \mathcal{T}$), if $\I$ satisfies each axiom in $\mathcal{R}$ (resp., $\mathcal{T}$) according to the semantic rules in Table \ref{semdlss}.  Analogously,  $\I= (\Delta^ \I, \Delta_{\D}, \cdot^\I)$ is a $\D$-model of $\mathcal{A}$, and we write $\I \models_{\D} \mathcal{A}$, if $\I$ satisfies each assertion in $\mathcal{A}$, according to Table \ref{semdlss}. A $\shdlssx$-KB $\mathcal{K}=(\mathcal{A}, \mathcal{T}, \mathcal{R})$ is consistent if there exists a $\D$-model $\I= (\Delta^ \I, \Delta_{\D}, \cdot^\I)$ of $\mathcal{A}$,  $\mathcal{T}$, and $\mathcal{R}$.


\subsubsection{The HOCQA problem for $\shdlssx$.}
We recall that the problem of \emph{Higher-Order Conjuctive Query Answering} (HOCQA) for $\shdlssx$, introduced in \cite{RR2017}, is a generalization of the Conjunctive Query Answering problem for $\shdlssx$ defined in \cite{ictcs16}. The HOCQA problem for $\shdlssx$ relies on the notion of \emph{Higher-Order} (HO) $\shdlssx$-conjunctive query,  admitting variables of three sorts: individual and data type variables, concept variables, and role variables. The HOCQA problem for $\shdlssx$ consists in finding the HO answer set of an HO-$\shdlssx$-conjunctive query (see below) with respect to a $\shdlssx$-KB.


Specifically, let $\varind  = \{\sfvar{v}{1}, \sfvar{v}{2}, \ldots\}$, 
$\vare = \{\sfvar{e_1}, \sfvar{e_2}, \ldots\}$,
$\vardt =\{\sfvar{t_1}, \sfvar{t_2}, \ldots \}$, 
$\varcon = \{\sfvar{c}{1}, \sfvar{c}{2}, \ldots\}$, 
$\varar = \{\sfvar{r}{1}, \sfvar{r}{2}, \ldots\}$,  
and $\varcr  = \{\sfvar{p}{1},$ $\sfvar{p}{2}, \ldots\}$ be pairwise disjoint denumerably infinite sets of variables 
 disjoint from $\Ind$, $\bigcup\{N_C(d): d \in N_{\D}\}$, $\C$, $\Ra$, and $\Rd$. HO-$\shdlssx$-\emph{atomic formulae} are expressions of the following types:
\[
R(w_1,w_2),
P(w_1, u),
C(w_1),
t(u),
\mathsf{r}(w_1,w_2),
\mathsf{p}(w_1, u),
\mathsf{c}(w_1),
\sfvar{t}{} (u),
w_1=w_2,
\] 
where $w_1,w_2 \in \varind \cup \Ind$, $u \in  \vare \cup \bigcup \{N_C(d): d \in N_{\D}\}$, $R$ is a $\shdlssx$-abstract role term, $P$ is a $\shdlssx$-concrete role term, $C$ is a $\shdlssx$-concept term,  $t$ is a $\shdlssx$-data type term, $\mathsf{r} \in \varar$, $\mathsf{p} \in \varcr$, $\mathsf{c} \in \varcon$, $\sfvar{t}{} \in \vardt$. A HO-$\shdlssx$-atomic formula containing no variables is said to be \emph{ground}. A HO-$\shdlssx$-\emph{literal} is a HO-$\shdlssx$-atomic formula or its negation. 
A HO-$\shdlssx$-\emph{conjunctive query} is a conjunction of HO-$\shdlssx$-literals. 
We denote with $\lambda$ the \emph{empty} HO-$\shdlssx$-conjunctive query.

Let  $\sfvar{v}{1},\ldots, \sfvar{v}{n} \in \varind$, $\sfvar{e}{1},\ldots,\sfvar{e}{g} \in \vare$, $\sfvar{t}{1},\ldots,\sfvar{t}{l} \in \vardt$,$\sfvar{c}{1}, \ldots, \sfvar{c}{m} \in \varcon$, $\sfvar{r}{1}, \ldots, \sfvar{r}{k} \in \varar$, $\sfvar{p}{1}, \ldots, \sfvar{p}{h} \in \varcr$, $o_1, \ldots, o_n \in \Ind$, $e_{d_1}, \ldots, e_{d_g} \in  \bigcup \{N_C(d): d \in N_{\D}\}$, $C_1, \ldots, C_m \in \C$, $R_1, \ldots, R_k \in \Ra$, and $P_1, \ldots, P_h \in \Rd$.
A substitution 	$\sigma  \defAs \{\sfvar{v}{1}/o_1, \ldots, \sfvar{v}{n}/o_n, \sfvar{e}{1}/{e_{d_1}}, \ldots, \sfvar{e}{g}/{e_{d_g}},  \sfvar{t}{1}/t_1, \ldots, \sfvar{t}{l}/t_l,\sfvar{c}{1}/{C_1}, \ldots, \sfvar{c}{m}/{C_m},\\ \null ~~\sfvar{r}{1}/{R_1}, \ldots, \sfvar{r}{k}/{R_k}, \sfvar{p}{1} /{P_1}, \ldots, \sfvar{p}{h}/{P_h} \}$ 
%
%
%
%
is a map such that, for every  HO-$\shdlssx$-literal $L$, $L\sigma$ is obtained from $L$ by replacing:
(a) the occurrences of $\sfvar{v}{i}$ in $L$ with $o_i$, for $i=1, \ldots, n$;
(b) the occurrences of $\sfvar{e}{b}$ in $L$ with $d_b$, for $b=1, \ldots, g$;
(c) the occurrences of $\sfvar{t}{s}$ in $L$ with $t_s$, for $s=1, \ldots, l$;
(d) the occurrences of $\sfvar{c}{j}$ in $L$ with $C_j$, for $j=1, \ldots, m$;
(e) the occurrences of $\sfvar{r}{\ell}$ in $L$ with $R_\ell$, for $\ell=1, \ldots, k$;
(f) the occurrences of $\sfvar{p}{t}$ in $L$ with $P_t$, for $t=1, \ldots, h$.
Substitutions can be extended to HO-$\shdlssx$-conjunctive queries in the usual way.

Let $Q \defAs  (L_1 \wedge \ldots \wedge L_m)$ be a HO-$\shdlssx$-conjunctive query, and $\KB$ a $\shdlssx$-KB. A substitution $\sigma$ involving \emph{exactly} the variables occurring in $Q$ is a \emph{solution for $Q$ w.r.t. $\KB$}, if there exists a $\shdlssx$-interpretation $\I$ such that $\I \models_{\D} \KB$ and $\I \models_{\D} Q \sigma$. The collection $\Sigma$ of the  solutions for $Q$ w.r.t. $\KB$ is the \emph{higher-order answer set of $Q$ w.r.t. $\KB$}. Then the \emph{higher-order conjunctive query answering problem} for $Q$ w.r.t. $\KB$ consists in finding the HO answer set $\Sigma$ of $Q$ w.r.t. $\KB$. 

The HOCQA problem for $\shdlssx$ can be instantiated to the most significant ABox reasoning problems for $\shdlssx$ (see \cite{RR2017}).

\subsubsection{Representing $\shdlssx$ in set-theoretic terms.}
$\shdlssx$-KBs and HO-$\shdlssx$-conjunctive queries can be represented in set-theoretic terms by exploiting a mapping $\theta$ defined in \cite{RR2017}. The function $\theta$ translates $\shdlssx$ statements in $\coreflqsr$-formulae in CNF. 
 Specifically, $\theta$ maps injectively individuals $a$, constants $e_d \in N_{C}(d)$, variables $w \in \varind$, and variables $u \in \vare$ into sort $0$ variables $x_a$, $x_{e_d}$, $x_{w}$, $x_{u}$, the constant concepts $\top$ and $\bot$, data type terms $t$, concept terms $C$, $\mathsf{c} \in \varcon$, and $\sfvar{t}{} \in \vardt$ into sort $1$ variables $X_{\top}^1$, $X_{\bot}^1$, $X_{t}^1$, $X_{C}^1$, $X_{\mathsf{c}}^1$, $X_{\sfvar{t}{}}^1$ respectively, and the universal relation 
 $U$, abstract role terms $R$, concrete role terms $P$, $\mathsf{r} \in \varar$, and $\mathsf{p} \in \varcr$ into sort $3$ variables $X_{U}^3$, $X_{R}^3$, $X_{P}^3$, $X_{\mathsf{r}}^3$, $X_{\mathsf{p}}^3$, respectively.\footnote{The use of level $3$ variables to model abstract and concrete role terms is motivated by the fact that their elements, that is ordered pairs $\langle x, y \rangle$, are encoded in Kuratowski's style as $\{\{x\}, \{x,y\}\}$, namely as collections of sets of objects.}

\noindent The mapping $\theta$  is defined for HO-$\shdlssx$-atomic formulae as follows:
%

\noindent $\theta (R (w_1,w_2)) \defAs \langle x_{w_1}, x_{w_2} \rangle \in X^3_{R}$,~~ $\theta (P (w_1,u)) \defAs \langle x_{w_1}, x_{u} \rangle \in X^3_{P}$,~~ $\theta ( C(w_1)) \defAs x_{w_1} \in X^1_{C}$,~~ $\theta ( t(u)) \defAs x_{u} \in X^1_{t}$,~~ $\theta (w_1 =w_2) \defAs x_{w_1} = x_{w_2}$,~~$\theta ( \sfvar{t}{}(u)) \defAs x_{u} \in X^1_{\sfvar{t}{}}$,~~$\theta (\sfvar{c}{} (w_1)) \defAs  x_{w_1} \in X^1_{\sfvar{c}{}}$,~~$\theta (\sfvar{r}{} (w_1,w_2)) \defAs \langle x_{w_1}, x_{w_2} \rangle \in X^3_{\sfvar{r}{}}$,~~$\theta (\sfvar{p}{} (w_1,u)) \defAs \langle x_{w_1}, x_{u} \rangle \in X^3_{\sfvar{p}{}}$.

\noindent Finally, $\theta$ is extended to HO-$\shdlssx$-conjunctive queries and to substitutions in a standard way. 
\begin{sloppypar}
From now on we denote with $\phi_{\KB}$ the $\coreflqsr$ translation of a $\shdlssx$-KB $\KB$ and with $\psi_Q$ the $\coreflqsr$-formula representing the HO-$\shdlssx$-conjunctive query $Q$. The formula $\phi_{\KB}$ is 
a conjunction of $\coreflqsr$-formulae of type $(\forall z_1) \ldots (\forall z_n) \varphi_0$, with $\varphi_0$ a clause of $\coreflqsr$-literals, since (a) each $\shdlssx$-KB $\KB$ is a set of statements $H$ such that  $\theta(H)$ is a $\coreflqsr$-formula in Table \ref{tablecore}; and (b)  $\phi_{\KB}$ is constructed by conjoining the $\theta(H)$s,  moving universal quantifiers as inward as possible, and renaming quantified variables as to be pairwise distinct. The interested reader is referred to \cite{RR2017ext} for full details.
\end{sloppypar}


Finally, the HOCQA problem for $\coreflqsr$-formulae can be stated as follows.
Let $\psi$ be a conjunction of $\coreflqsr$-literals and $\phi$ a $\coreflqsr$-formula. The \emph{HOCQA problem for $\psi$ w.r.t.\ $\phi$} consists in computing the HO \emph{answer set of $\psi$ w.r.t.\ $\phi$}, namely the collection $\Sigma'$ of all the  substitutions $\sigma'$ such that  $\M \models \phi \wedge \psi\sigma'$, for some $\coreflqsr$-interpretation $\M$.

\section{Overview of the reasoner}
We present a general overview  of the reasoner and the main notions and definitions concerning the procedures upon which it is based. 

\begin{figure}[h] 
	\centering
	\includegraphics[scale=0.5]{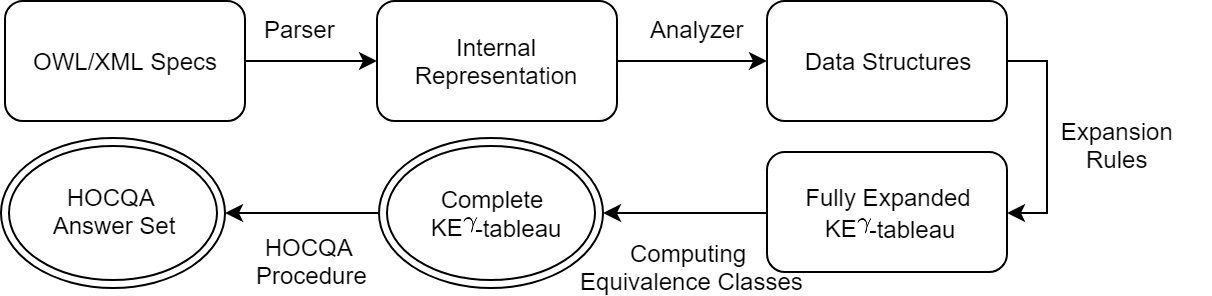} \caption{Execution cycle of the reasoner.} \label{FigOver}
\end{figure}

The input of the reasoner is an OWL ontology serialized in the OWL/XML syntax and admitting SWRL rules (see Figure \ref{FigOver}). 
\vspace*{-0.5cm}

If the ontology meets the $\shdlssx$ requirements, a parser produces the internal coding of all axioms and assertions of the ontology in set-theoretic terms, as a list of strings. Then the system builds the data-structures required to execute the algorithm. In the two subsequent steps, the reasoner constructs a complete \keg\space $\T_\KB$ whose open branches represent all possible models for the input KB $\phi_\KB$ (see below for the definition of \keg). The tableau $\T_\KB$ is constructed (1) by systematically applying the following two rules: (1a) a generalization of the KE-elimination rule incorporating the $\gamma$-rule, and (1b) the principle of bivalence rule (PB-rule) (thus constructing all branches of the \keg---see Figure \ref{exprule}), and then (2) processing each open branch $\vartheta$ of  $\T_\KB$ by constructing the equivalence classes of the individuals involved in formulae of type $x=y$ occurring in $\vartheta$ and substituting each individual $x$ on $\vartheta$ with the representative of the equivalence class of $x$. Such step returns the complete \keg. Finally, the reasoner takes as input the internal coding of $\psi_Q$, i.e. the set-theoretic representation of a query $Q$, and computes the HO-answer set of $\psi_Q$ with respect to $\phi_\KB$.
The task of computing the complete \keg\space for $\phi_{\KB}$ is performed by procedure $\consistency$ illustrated in Figure \ref{proc1}, whereas the task of computing the HOCQA answer set of a given query w.r.t the KB is performed by procedure $\prochoplus$ shown in Figure \ref{proc2}. 

\begin{figure}[H]
	{\scriptsize
		\begin{algorithmic}[1]
			\Procedure{$\consistency$}{$\phi_\KB$}
			\State $\Phi_\KB :=\{\phi: \phi \mbox{ is a conjunct of } \phi_{\KB}\}$
			\State $\T_{\KB}$ := $\Phi_\KB$;
			\State $\mathcal{E}:= \emptyset$;
			\While{$\T_{\KB}$ is not fulfilled}
			\parState{- select a not fulfilled open branch $\vartheta$ of $\T_{\KB}$ and a not fulfilled formula\\ \hspace*{.2cm}$\psi=(\forall{x_1})\ldots(\forall{x_m})(\beta_1 \vee \ldots \vee \beta_n)$ in $\vartheta$;}
			%
			%
			%
			%
			%
			\parState{ $\Sigma^{\KB}_\psi = \{\tau : \tau=\{x_1/x_{o_m}, \ldots, x_m/x_{o_m}\}\}$, where $\{x_1,\ldots,x_m\}=Q\var_0(\psi)$ and $\{x_{o_1},\ldots,x_{o_m}\}\in \varz(\phi_{\KB})$;}
			\For{$\tau \in \Sigma^{\KB}_\psi$} 
			\If{$\beta_i\tau \notin \vartheta$, for every $i=1,\ldots,n$}
			\If{$\seqsnj$ is in $\vartheta$, for some $j \in \{1,\ldots,n\}$}
			\State - apply the $\egamma$ to $\psi$ and $\seqsnj$ on $\vartheta$; \label{procCon:Egamma}
			\Else 
			\parState{- let $B^{\overline{\beta}\tau}$ be the collection of literals $\overline{\beta}_1\tau,\ldots,\overline{\beta}_n\tau$ present in $\vartheta$
				and let\\ \hspace*{.15cm} $h$ be the lowest index such that $\overline{\beta}_h\tau \notin B^{\overline{\beta}\tau}$;}
			\parState{- apply the PB-rule to $\overline{\beta}_h\tau$ on $\vartheta$;} \label{procCon:pbrule}
			\EndIf;
			\EndIf;
			\EndFor;
			\EndWhile;
			\For {$\vartheta$ in $\T_{\KB}$}
			\If {$\vartheta$ is an open branch}
			\label{procCon:whileBranch}
			%
			
			\State $\sigma_{\vartheta} := \epsilon$ (where $\epsilon$ is the empty substitution);

			\State $\mathsf{Eq}_{\vartheta} := \{ \mbox{literals of type $x = y$, occurring in $\vartheta$}\}$;
			
			\While{$\mathsf{Eq}_{\vartheta}$ contains $x = y$, with distinct $x$, $y$} \label{procCon:whileEQ}
			
			\State - select a literal $x = y$ in $\mathsf{Eq}_{\vartheta}$, with distinct $x$, $y$;
			
			\State $z :=$ $min_{\minord}(x,y)$;
			
			\State $\sigma_{\vartheta} := \sigma_{\vartheta} \cdot \{x/z, y/z\}$;
			
			\State $\mathsf{Eq}_{\vartheta} := \mathsf{Eq}_{\vartheta}\sigma_{\vartheta}$;
			\EndWhile;
			
			\State $\mathcal{E} = {\mathcal{E} \cup \{(\vartheta,\sigma_\vartheta)\}}$;
			\State $\vartheta := \vartheta\sigma_{\vartheta}$;
			\EndIf;
			\EndFor;
			\State \Return $(\T_\KB, \mathcal{E})$;
			\EndProcedure;
		\end{algorithmic}
	}
	\caption{The procedure $\consistency$.}
	\label{proc1}
\end{figure}

\vspace*{1cm}
\begin{figure}
	{\small
		\begin{algorithmic}[1]
			\Procedure{$\prochoplus$}{$\psi_Q$, $\mathcal{E}$}
			
			\State $\Sigma'$ := $\emptyset$;	
			\parWhile{$\mathcal{E}$ is not empty}
			\State - let  $(\vartheta,\sigma_\vartheta) \in \mathcal{E}$;
			\State - $\vartheta := \vartheta\sigma_{\vartheta}$;
			\State - initialize $\mathcal{S}$ to the empty stack;
			
			\State - push $(\epsilon, \psi_Q\sigma_\vartheta)$ in $\mathcal{S}$;
			
			\While{$\mathcal{S}$ is not empty}
			\State - pop $(\sigma', \psi_Q\sigma_\vartheta\sigma')$ from $\mathcal{S}$;
			
			\If{$\psi_Q\sigma_\vartheta\sigma' \neq \lambda$}
			\State - let $q$ be the leftmost conjunct of $\psi_Q\sigma_\vartheta\sigma'$;
			
			\State $\psi_Q\sigma_\vartheta\sigma':= \psi_Q\sigma_\vartheta\sigma'$ deprived of $q$;
			\State $\litqt := \{ t \in \vartheta : t=q\rho$, for some substitution $\rho \}$; \label{procHOP:compLit}
			
			\While{$\litqt$ is not empty} \label{procHOP:whileLit}
			\State - let $t \in \litqt$, $t=q\rho$;
			
			\State $\litqt := \litqt \setminus \{t\}$;
			
			\State - push $(\sigma'\rho, \psi_Q\sigma_\vartheta\sigma'\rho)$ in $\mathcal{S}$;
			\EndWhile;
			\Else
			\State $\Sigma'$ := $\Sigma' \cup \{\sigma_\vartheta\sigma'\}$;
			\EndIf;
			\EndWhile;
			\State $\mathcal{E} := {\mathcal{E} \setminus \{(\vartheta,\sigma_\vartheta)\}}$; \label{procHOP:Edel}
			\EndparWhile;	
			\State \Return $\Sigma'$;
			\EndProcedure;
		\end{algorithmic}
	}
	\caption{The procedure $\prochoplus$.}
	\label{proc2}
\end{figure}


 Let $\Phi_\KB \defAs \{\phi: \phi \mbox{ is a conjunct of } \phi_{\KB}\}$. The procedure $\consistency$ constructs a complete  \keg\space $\T_{\KB}$ for the set $\Phi_\KB$ of the conjuncts of $\phi_{\KB}$ representing the saturation of the $\shdlssx$-KB.  
 
 At this point, it is convenient to give the definition of \kegx.
 Let $\Phi \defAs \{ C_1,\ldots, C_p\}$, where each $C_i$ is either a $\coreflqsr$-literal of the types in Table \ref{tablecore} or a $\coreflqsr$-purely universal quantified formula of the form $(\forall{x_1})\ldots(\forall{x_m})(\beta_1 \vee \ldots \vee \beta_n)$, with $\beta_1\ldots \beta_n$ $\coreflqsr$-literals. $\mathcal{T}$ is a \emph{\keg} for $\Phi$ if there exists a finite sequence $\mathcal{T}_1, \ldots, \mathcal{T}_t$ such that (i) $\mathcal{T}_1$ is a one-branch tree consisting of the sequence $C_1,\ldots, C_p$, (ii) $\mathcal{T}_t = \mathcal{T}$, and (iii) for each $i<t$, $\mathcal{T}_{i+1}$ is obtained from $\mathcal{T}_i$ either by an application of one of the rules ($\egamma$ or PB-rule) in Figure \ref{exprule} or by applying a substitution $\sigma$ to a branch $\vartheta$ of $\mathcal{T}_i$ (in particular, the substitution $\sigma$ is applied to each formula $X$ of $\vartheta$ and the resulting branch will be denoted with $\vartheta\sigma$). In the definition of the $\egamma$ reported in Figure \ref{exprule}, (a) $\tau :=\{x_1/x_{o_1} \ldots x_m/x_{o_m}\}$ is a substitution such that $x_1,\ldots,x_m$ are the quantified variables in $\psi$ and $x_{o_1}, \ldots, x_{o_m} \in \varz(\phi_{\KB})$; and (b) $\seqs \defAs \{ \overline{\beta}_1\tau,\ldots,\overline{\beta}_n\tau\} \setminus \{\overline{\beta}_i\tau\}$ is a set containing the  complements of all the disjuncts $\beta_1 \ldots \beta_n$ to which the substitution $\tau$ is applied, with the exception of the disjunct $\beta_i$.

Initially, the procedure $\consistency$  constructs a one-branch \keg\space $\T_{\KB}$ for the set $\Phi_\KB$ of conjuncts of $\phi_\KB$. Then, it expands $\T_{\KB}$ by systematically applying the $\egamma$  and the PB-rule  in Figure \ref{exprule} to formulae of type $\psi=(\forall{x_1})\ldots(\forall{x_m})(\beta_1 \vee \ldots \vee \beta_n)$ till they are all fulfilled, giving priority to 
the $\egamma$. Once such rules are no longer applicable, for each open branch $\vartheta$ of the resulting \keg, atomic formulae of type $x = y$ occurring in $\vartheta$ are used to compute 
the equivalence class of $x$ and $y$.  For each open branch $\vartheta$ of $\T_{\KB}$, the equivalence class of each variable occurring in $\vartheta$ is obtained by computing the  substitution $\sigma_\vartheta$ such that $\vartheta\sigma_\vartheta$ does not contain literals of type $x=y$, for distinct $x,y$. The resulting pair $(\vartheta, \sigma_\vartheta)$ is added to the set $\mathcal{E}$.  

\begin{figure}
	{{\footnotesize
			\begin{center}
				\begin{minipage}[h]{5cm}
					$\infer[\textbf{E}^{\mathbf{\gamma}}\textbf{-rule}]
					{\beta_i\tau}{\psi & \quad \seqs}$\\[.1cm]
					{ where\\[-.1cm] $\psi \defAs (\forall{x_1})\ldots(\forall{x_m})(\beta_1 \vee \ldots \vee \beta_n)$,\\
						$\tau :=\{x_1/x_{o_1} \ldots x_m/x_{o_m}\}$,\\
						and $\seqs \defAs \{ \overline{\beta}_1\tau,...,\overline{\beta}_n\tau\} \setminus \{\overline{\beta}_i\tau\}$,}{ for $i=1,...,n$}
				\end{minipage}~~~~~~~~~~~
				\begin{minipage}[h]{3.5cm}
					\vspace{-1.12cm}
					$\infer[\textbf{PB-rule}]
					{A~~|~~\overline{A}}{}$\\[.1cm]
					{ where $A$ is a literal}
				\end{minipage}
			\end{center}
			\vspace{-.2cm}
		}
		\caption{\label{exprule} Expansion rules for the \keg.}
	}
\end{figure}
\vspace*{-0.3cm}

The procedure $\prochoplus$ takes as input a query $\psi_Q$ and the set $\mathcal{E}$ yielded by the procedure $\consistency$ and returns the answer set $\Sigma'$ of $\psi_Q$ w.r.t. $\phi_\KB$. For each open and complete branch $\vartheta$ of $\T_\KB$, the procedure $\prochoplus$ builds a decision tree $\DT_{\vartheta}$ where each  maximal branch induces a substitution $\sigma'$ such that $\sigma_\vartheta\sigma'$ belongs to the answer set of $\psi_Q$ w.r.t. to $\phi_\KB$.

$\DT_{\vartheta}$ is obtained by constructing a stack of its nodes. Initially the stack contains just the root node $(\epsilon,\psi_Q\sigma_\vartheta)$ of  $\DT_{\vartheta}$, with $\epsilon$ the empty substitution. At each step, the procedure pops out from the stack an element $(\sigma', \psi_Q\sigma_\vartheta\sigma')$  and iteratively selects a literal $q$ from the query $\psi_Q\sigma_\vartheta\sigma'$ and eliminates it from $\psi_Q\sigma_\vartheta\sigma'$.  Then, the set of literals $t$ in $\vartheta$ matching $q$ is computed by putting $\litqt := \{ t \in \vartheta : t=q\rho$, for some substitution $\rho \}$.  The successors of the current node are computed by pushing the node $(\sigma'\rho, \psi_Q\sigma_\vartheta\sigma'\rho)$ in the stack, for each element in $\litqt$ . If the current node  has the form of $(\sigma',\lambda)$, with $\lambda$ the empty query, the last literal of $\psi_Q$ has been treated and the substitution $\sigma_\vartheta\sigma'$  is inserted in $\Sigma'$. Notice that, in case of a failing query match, the set $\litqt$  is empty and then no successor node is pushed into the stack. Thus, the failing branch of $\DT_{\vartheta}$ is abandoned and  another branch is selected  by popping  one of the nodes of $\DT_{\vartheta}$ from the stack. 

Computational complexity results can be found in \cite{RR2018ext}.

\subsection{Some implementation details}

We first show how the internal coding of $\shdlssx$-KBs is represented in terms of $\coreflqsr$-formulae and the data-structures used by the reasoner for representing formulae, nodes, and how \kegx\ are implemented. Then we describe the most relevant functions that implement the procedures $\consistency$ and $\prochoplus$ and also illustrate an example of reasoning in $\shdlssx$. 

To begin with, $\coreflqsr$-variables, quantifiers, Boolean operators, set-theoretic relators, and pairs are mapped into strings as follows. Variables of type $X^i_{\mathit{name}}$ are mapped into strings of the form $\mathit{Vi}\{\mathit{name}\}$. For the sake of uniformity, variables of sort 0 are denoted with $X^0, Y^0, \ldots$, whereas individuals $a$,  concepts $C$, and roles $R$ of a $\shdlssx$-KB are respectively mapped into the variables $X^0_a$, $X^1_C$, and $X^3_R$, according to the function $\theta$ described in \cite{RR2017}. 
The symbols $\forall$, $\wedge$, $\vee$, $\neg \wedge$, $\neg \vee$ are mapped into the strings \texttt{\$FA}, \texttt{\$AD}, \texttt{\$OR}, \texttt{\$DA}, \texttt{\$RO}, respectively. The relators $\in$, $\not \in$, $=$, $\neq$ are mapped into the strings \texttt{\$IN}, \texttt{\$NI}, \texttt{\$EQ}, \texttt{\$QE}, respectively. A pair $\langle X^0_1,X^0_2\rangle$ is mapped into the string \texttt{ \$OA\, V0{1}\, \$CO \,V0{2}\, \$AO}, where \texttt{\$OA} represents the bracket \textquotedblleft $\langle$'', \texttt{\$AO} the bracket \textquotedblleft {$\rangle$}", and \texttt{\$CO} the comma symbol.

Then, data-structures for representing the KB are built.
$\coreflqsr$-variables are implemented by means of the class \texttt{Var} that has four fields. The field \texttt{type} of type integer  indicates the sort of the variable, the field \texttt{name} of type string represents the name of the variable, and  the field \texttt{var}  of type integer represents a free variable if set to 0, and a quantified 
variable if set to 1. The field \texttt{index} stores the position of the variable in the vector \texttt{VVL}, delegated to collect free variables. Quantified and free variables are collected  in the vectors  \texttt{VQL} and \texttt{VVL} respectively, which provide a subvector for each sort of variable. 

The operators admitted in $\coreflqsr$, internally coded as strings, are mapped into three vectors that are fields of the class \texttt{Operator}. Specifically, the vector  \texttt{boolOp}  contains the values \texttt{\$OR}, \texttt{\$AD}, \texttt{\$RO}, \texttt{\$DA},  the vector \texttt{setOp} the values \texttt{\$IN}, \texttt{\$EQ}, \texttt{\$NI}, \texttt{\$QE}, \texttt{\$OA}, \texttt{\$AO}, \texttt{\$CO}, and the vector \texttt{qutOp} the value \texttt{\$FA}. 

%
%

$\coreflqsr$-literals are stored using the class \texttt{Lit} that has two fields. The field \texttt{litOp} of type integer represents the operator of the formula and corresponds to the index of one of the first four elements of the vector \texttt{setOp}. The field \texttt{components} is a vector whose elements point to the variables 
involved in the literal and stored in \texttt{VQL} and \texttt{VVL}. 


$\coreflqsr$-formulae are represented by the class \texttt{Formula}, having a binary tree structure, whose nodes contain objects of the class \texttt{Lit}.
The left and right children contain the left and right subformula, respectively. The class \texttt{Formula} contains the following fields: the field \texttt{lit} of type pointer to \texttt{Lit} represents the literal; the field \texttt{operand} represents the propositional operator, and its value is the index of the corresponding element of the vector \texttt{boolOp}; 
the field \texttt{psubformula} of type pointer to \texttt{Formula} is the pointer to the father node, whereas the fields \texttt{lsubformula} and \texttt{rsubformula} contain the pointers to the
nodes representing the left and the right component  of the formula, respectively.


The procedure $\consistency$ is based on the data-structure implemented by the class \texttt{Tableau}. This class uses the instances of the class \texttt{Node} that represents the nodes of the \keg. The class \texttt{Node} has a tree-shaped structure and four fields:  the field \texttt{setFormula}, of type vector of \texttt{Formula}, that collects the formulae of the current node, and three pointers to instances of the class \texttt{Node}. These are the \texttt{leftchild}, \texttt{rightchild}, and \texttt{father} fields, which point to the left child node, right child node, and father node, respectively. 

The root node of  the class \texttt{Tableau} contains the field \texttt{root} of type pointer to \texttt{Node}. The fields  \texttt{openbranches} and \texttt{closedbranches} collect the set of open branches and of closed branches, respectively.
 In addition, the class \texttt{Tableau} is provided with the field \texttt{EqSet} that is a three-dimensional vector of integers storing the equivalence classes induced by atomic formulae of type $X^0 = Y^0$. 
In particular, \texttt{EqSet} stores 
a vector containing the indices of \texttt{VVL} corresponding to the variables belonging to the equivalence classes.

%
%
 
As mentioned above, the reasoner takes as input  an OWL ontology compatible with the $\shdlssx$ requirements, also admitting SWRL rules, and serialized in the OWL/XML syntax. As first step, the function \texttt{readOWLXMLontology} produces the internal coding of all axioms and assertions of the ontology, yielding a list of strings. Then the reasoner builds from the output of \texttt{readOWLXMLontology} the objects of type \texttt{Formula} that implement the $\coreflqsr$-formulae representing the KB, and stores them in the field \texttt{root} of an object of type \texttt{Tableau}. In this phase, formulae are transformed in CNF and universal quantifiers are moved as inward as possible and renamed in such a way as to be pairwise distinct.
 The  object of type \texttt{Tableau} representing the \keg\space is the input to the procedure \texttt{expandGammaTableau} that expands the \keg\space by iteratively selecting and fulfilling purely universal quantified input formulae. Once a purely universal quantified formula has been selected, \texttt{expandGammaTableau} builds iteratively the set of substitutions $\tau$ to be applied to the selected formula.  A substitution $\tau$ is a map from the indices of the quantified variables of the formula, selected in order of appearance, to the elements of the vector \texttt{VVL}. 
  The implementation of $\tau$  applies standard techniques for computing the \textit{variations with repetition} of the set of indices of the elements of \texttt{VVL} taken to $k$ by $k$, where $k$ is the number of quantified variables occurring in the selected formula.
 
 The procedure \texttt{expandGammaTableau} fulfills the formula selected by systematically applying the functions \texttt{EGrule} with the current $\tau$ and \texttt{PBrule}, respectively implementing the $\egamma$ and the PB-rule. More precisely, it works as follows. The disjuncts of the current formula to which $\tau$ is applied are stored in a temporary vector and selected iteratively. If a disjunct has its negation on the branch, it is removed from the temporary vector. Once all the elements of the temporary vector have been selected, if the last one does not have its negation on the branch, then \texttt{EGrule} is applied to the formula and the last element of the temporary vector is inserted in the branch according to Figure \ref{exprule}. If there is more than one element left in the temporary vector, then the procedure \texttt{PBrule} is applied. In case the stack is empty, a contradiction is found and the branch gets closed and inserted in the vector \texttt{closedbranches}.

If the procedure \texttt{expandGammaTableau} terminates with some elements in \texttt{openbranches}, then the reasoner builds the set of equivalence classes of the variables involved in formulae of type $X^0=Y^0$, for each element of \texttt{openbranches} by means of the procedure \texttt{buildsEqSet}. The latter procedure updates the field \texttt{EqSet} of the object of type \texttt{Tableau} with the new information concerning the set of equivalence classes. After the execution of \texttt{buildsEqSet}, if \texttt{openbranches} contains some elements, a consistent KB is returned.

Procedure $\prochoplus$ is implemented by the function $\mathtt{performQuery}$ that takes as input  the object of type \texttt{Tableau} returned by \texttt{buildsEqSet} and a string representing the internal coding of the input query $\psi_Q$, and returns  an object of type $\mathtt{QueryManager}$ storing, among other information, the answer set of $\psi_Q$ w.r.t. $\phi_{\KB}$.
The function $\mathtt{performQuery}$ uses an object of type $\texttt{QueryManager}$ that stores the input query $\psi_Q$ as a string, an object of type \texttt{Formula} representing $\psi_Q$, and the answer set of $\psi_Q$ w.r.t. $\phi_{\KB}$, for each element of \texttt{openbranches}.  The answer set is implemented by endowing the object of type  $\mathtt{QueryManager}$ with the pair of vectors \texttt{VarMatch}. The first vector of \texttt{VarMatch} contains an integer for each element in \texttt{openbranches}: this is set to $1$ if the corresponding branch has solution, $0$ otherwise. 
 The second (three-dimensional) vector contains for each element in \texttt{openbranches} 
 a vector of solutions, each one constituted by a vector of pairs of pointers to \texttt{Var}. The first \texttt{Var} of such pair is a variable belonging to the query, whereas the second \texttt{Var} is the matched individual.

For each element in \texttt{openbranches}, the function $\texttt{performQuery}$ implements a decision tree by means of a stack that keeps track of the partial solutions of the query, as nodes of the decision tree. Such a stack, called \texttt{matchSet}, is constituted by a vector of pairs of objects of type \texttt{Var} such that the first one represents the query variable and the second one the matched element. Initially, \texttt{matchSet} is empty. At first step, the procedure selects the first conjunct of the query and, for each match found, it pushes in \texttt{matchSet} a vector of pairs representing the match. The procedure selects iteratively the conjuncts of the query and then applies to the selected conjunct the substitution that is currently at the top of \texttt{matchSet}. If the literal obtained by the application of such partial solution has one or more matches in the branch, the resulting substitutions are pushed in \texttt{matchSet}. Once all the literals of the query have been processed, if  \texttt{matchSet} is not empty, it contains the leaves of the maximal branches of the decision tree, which are all added to \texttt{VarMatch}. 

\subsection{Example of reasoning in $\shdlssx$}

\begin{figure}\centering
	\includegraphics[scale=0.7]{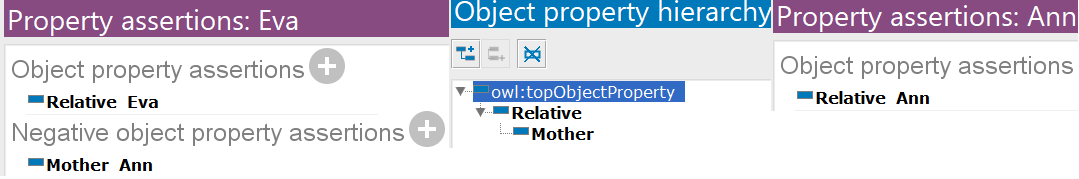}  	
	\caption{OWL ontology of the example.} \label{imgOWL}
\end{figure}

Let us consider the ontology displayed in Figure \ref{imgOWL}. 
\vspace*{-0.6cm}
\vspace*{-0.6cm}

The KB in terms of $\coreflqsr$ is the following formula:

\[ \footnotesize
\begin{array}{rcl}
\phi_{\KB} & \defAs & \neg ( \langle x_{Eva}, x_{Ann} \rangle \in X^3_{Mother} ) \wedge {} \\ &&\langle x_{Ann}, x_{Ann} \rangle \in X^3_{Relative} \wedge \langle x_{Eva}, x_{Eva} \rangle \in X^3_{Relative} \wedge {}\\ &&(\forall z_1)(\forall z_2)(  \neg ( \langle z_1, z_2 \rangle \in X^3_{Mother}) \vee \langle z_1, z_2 \rangle \in X^3_{Relative})
\end{array}
\]

\begin{figure}[H] 
	\centering \includegraphics[scale=0.4]{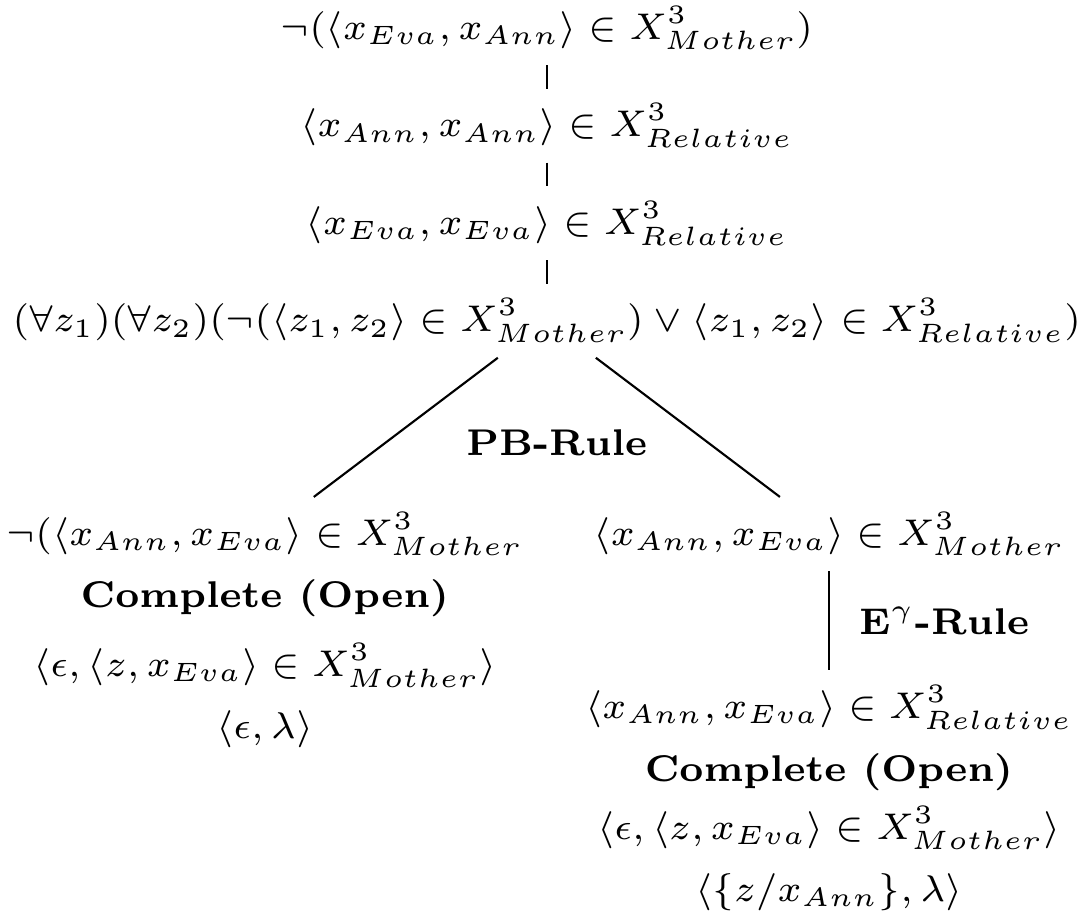}
	\caption{\keg\space for $\phi_{\KB}$ and decision trees for the evaluation of $\psi_Q$.}
	\label{kegfig}
\end{figure}

%
%
Let $\psi_Q= \langle z, x_{Eva} \rangle \in X^3_{Mother}$ be a query represented in set-theoretic terms. A complete \keg\space for $\phi_{\KB}$ and the decision trees constructed for the evaluation of $\psi_Q$ on each open branch of the \keg\space are shown in Figure \ref{kegfig}. Notice that, the decision tree constructed on the leftmost open branch of the \keg\space provides no solution.

\begin{figure}[H]\centering
	\includegraphics[scale=0.7]{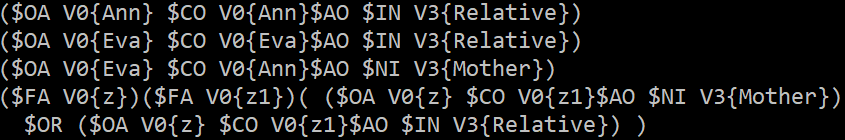}  	
	\caption{Internal representation of the ontology.}
	\label{intcode}
\end{figure}

\begin{figure}[H]\centering
 	\includegraphics[scale=0.6]{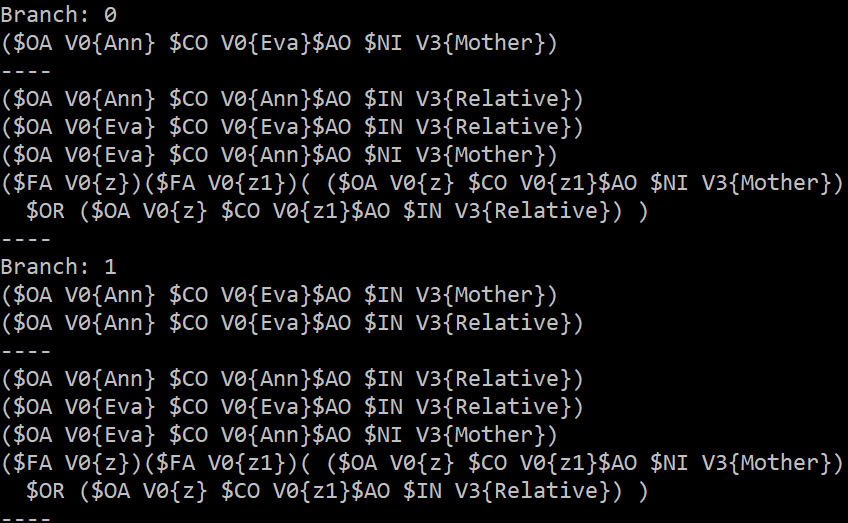}  
 	\includegraphics[scale=0.7]{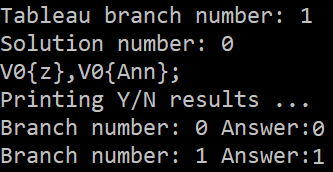} 	
 	\caption{The \keg\space and the evaluation of $\psi_Q$ computed by the reasoner.}
 	\label{kegtab}
 \end{figure}

\begin{figure}[H] 
	\centering \includegraphics[scale=0.32]{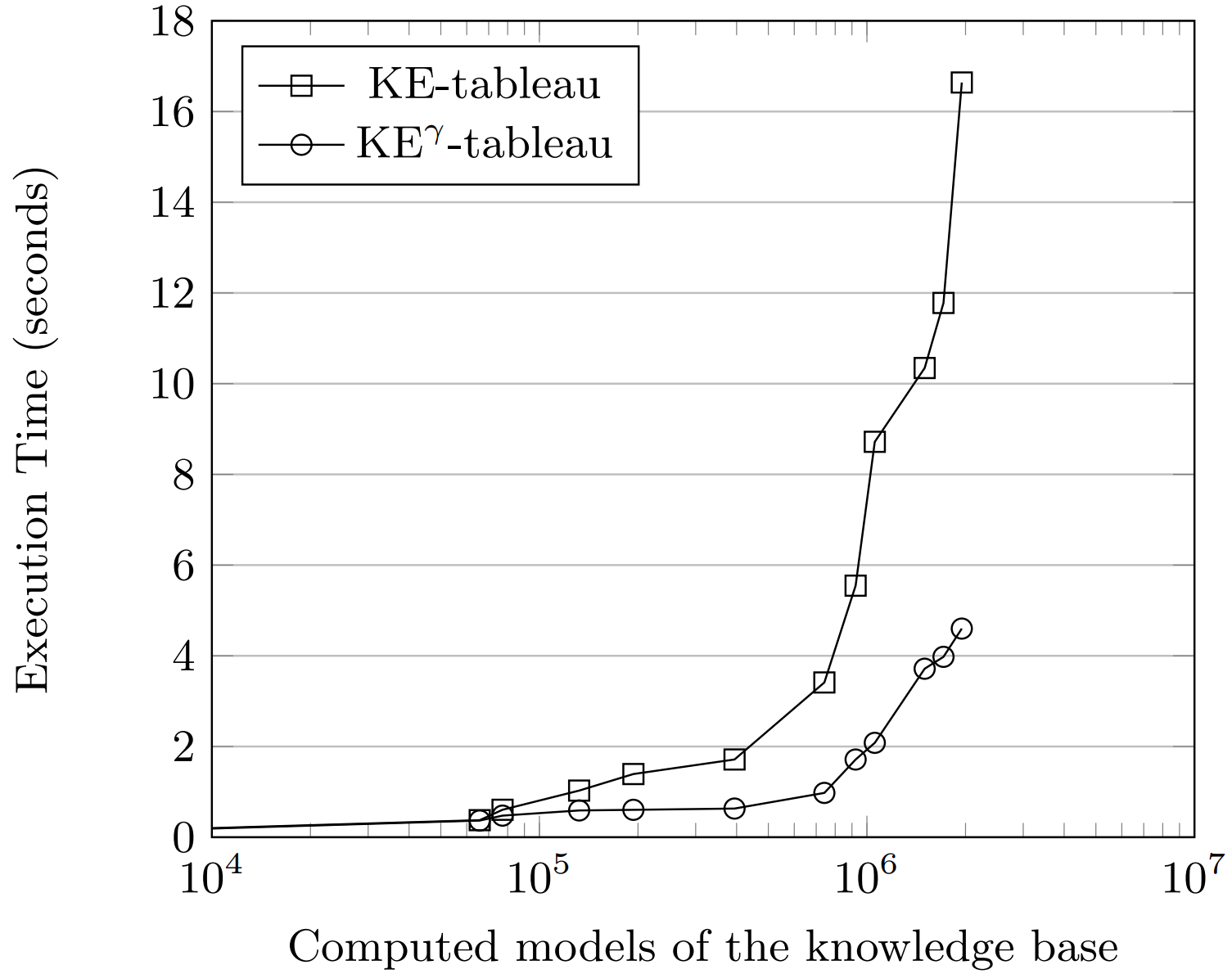}
	\caption{Comparison between \ke\space and \keg\space systems.}	\label{bench}
\end{figure}

The internal representation of the OWL ontology  is shown in Figure \ref{intcode}. The \keg\space computed by the reasoner and the evaluation of the query $\psi_Q$ are reported in Figure \ref{kegtab}.
Finally, Figure \ref{bench} shows a performance comparison between our implementation of the \ke\space presented in \cite{cilc17} and the \keg\space system for $\shdlssx$ presented in this paper. The metric used in the benchmarking is the number of models of the input KB computed by the reasoners and the time required to compute such models.  As shown in Figure \ref{bench}, the \keg\space has a better performance than the \ke\ up to about $400\%$, even if in some cases the performances of the two systems are comparable, as shown in the plot.  We conclude that the \keg\space system is always convenient, also because the expansions of quantified formulae are not stored in memory.

\vspace{-.53cm}

\vspace{-1cm}

\vspace*{-.8cm}
\vspace*{-0.8cm}

\section{Conclusions}
We presented a C++ implementation of a \keg\space system for the most widespread reasoning tasks of $\shdlssx$, such as consistency checking of $\shdlssx$-KBs and a generalization of the CQA problem for $\shdlssx$, , called HOCQA problem, admitting conjunctive queries with variables of three sorts. These problems have been addressed by translating $\shdlssx$-KBs and higher-order $\shdlssx$-conjunctive queries in terms of formulae of the set-theoretic fragment $\coreflqsr$. The reasoner is an improvement of the \ke\space system introduced in \cite{cilc17} to check consistency of $\shdlssx$-KBs, as it admits a generalization of the KE-elimination rule incorporating the $\gamma$-rule. 
The reasoner takes as input OWL ontologies compatible with the specifications of $\shdlssx$ serialized in the OWL/XML format and admitting SWRL rules.
 
Finally, we showed that the reasoner presented in this paper is more efficient than the one introduced in \cite{cilc17}, by means of suitable benchmark test sets. 

We plan to extend the set-theoretic fragment underpinning the reasoner to include also a restricted version of the operator of relational composition. This will allow ones to reason with description logics that admit full existential and universal quantification. In addition, we intend to improve our reasoner so as to deal with the reasoning problem of ontology classification. Then, we shall compare the resulting reasoner with existing well-known reasoners such as HermiT \cite{ghmsw14HermiT} and Pellet \cite{PelletSirinPGKK07}, providing some benchmarking. We also plan to allow data type reasoning by either integrating existing solvers for the Satisfiability Modulo Theories (SMT) problem or by designing ad-hoc new solvers. 
 Finally, as each branch of the \keg\space can be computed by a single processing unit, we plan to implement a parallel version of the software by using the Nvidia CUDA library.

\bibliographystyle{plain}
\bibliography{biblio}
\end{document}